# *In situ* Study of p-NiO Film Quality at High-Temperatures up to 1100 °C

Hunter Ellis[1,*], Bobby G. Duersch[2], Botong Li[1], Imteaz Rahaman[1], Jim Pierce[3], Michael A. Scarpulla[1,4], Kai Fu[1,*]

[1]Department of Electrical and Computer Engineering, The University of Utah, Salt Lake City, UT 84112, USA

[2]Utah Nanofab Electron Microscopy and Surface Analysis Laboratory, The University of Utah, Salt Lake City, UT 84112, USA

[3]Utah Nanofab, Price College of Engineering, The University of Utah, Salt Lake City, UT 84112, USA

[4]Department of Materials Science and Engineering, The University of Utah, Salt Lake City, Utah 84112, USA

[*]E-mail: u0973796@utah.edu; kai.fu@utah.edu

**Abstract:** NiO is a promising p-type material for photovoltaics and power electronics, but its temperature limits are unclear. Using *in situ* HT-XRD from 30 to 1100 °C, we track the evolution of the NiO film in air: it crystallizes from amorphous to cubic NiO between 300 and 400 °C, with the growth of the (111) peak matching the rising resistivity. Higher temperatures increase crystallinity and resistivity. At 1100 °C, $Ni_2O_3$ forms, producing a highly resistive film. The study presents a clear evolution of NiO film crystallinity in correlation with the resistive behavior.

Cubic nickel oxide (NiO) is a promising material for heterojunction devices. Its wide bandgap of 4.3 eV,[1] and its capability for reliable p-type doping through nickel vacancies makes NiO one of the few viable wide-bandgap p-type oxides available.[2] NiO can be deposited by RF sputtering in a mixed Ar/$O_2$ ambient, where oxygen incorporation facilitates the formation of Ni vacancies that act as acceptors. Due to the ease of deposition with controllable hole concentration, NiO is commonly incorporated in photovoltaics as a hole transport layer[3] and as a p-type layer in heterojunction diodes with GaN and ZnO.[4,5] NiO has also been used successfully for heterojunction devices with β-$Ga_2O_3$, with diodes demonstrating breakdown voltages greater than 8 kV at room temperature,[6-8] and stable breakdown voltages at temperatures up to 300 °C,[9] along with FETs,[10,11] and detectors.[12-16] Despite these advantages, NiO remains susceptible to degradation at elevated temperatures due to changes in crystal quality and reductions of Ni vacancies.[17,18] These effects can destabilize the physical and electronic properties of the film and the performance of devices incorporating NiO.

Previous studies have investigated the influence of post-deposition annealing on the crystallinity and electrical properties of NiO films.[19-23] However, such measurements only probe the final structural state after thermal exposure. For applications in high-temperature electronics, an *in situ* analysis is needed to understand the phase evolution and reliability of the material under the operating temperature to quantify the limits of NiO for high-temperature device applications. This work addresses that gap by performing *in situ* high-temperature X-ray diffraction (HT-XRD) measurements from 30 °C to 1100 °C on three different sputtered NiO films.

NiO films were deposited onto c-plane sapphire using RF sputtering from a 99.9% pure NiO target. The base pressure prior to deposition was $8 \times 10^{-6}$ mTorr. Ar and $O_2$ were introduced at equal flow rates of 6 sccm to the sputtering chamber. The films were deposited at sputtering pressures of 2.5, 5.0, and 10 mTorr using 50 W RF power, producing approximately 40 nm of NiO in each case. Structural characterization was performed using a Bruker D8 Discover diffractometer. The XRD was equipped with an Eiger 2R 250k detector and a 0.20 mm slit. An Anton Paar DHS 1100 heating stage was used to heat the samples, and a graphite dome enclosed the sample, protecting the instrumentation during high-temperature operation.

*In situ* XRD scans were conducted from 30 °C to 1100 °C, with a 2θ range of 18.5° to 45°. This measurement was followed by a post-high-temperature scan at 30 °C. The temperature was increased at a rate of 100 °C min$^{-1}$ with a 2-min dwell at each set point. After reaching 1100 °C, each sample was cooled to room temperature in approximately 30 min.

Figures 1(a)–1(c) show the evolution of the NiO films deposited at 2.5, 5.0, and 10 mTorr. At temperatures up to ~200 °C, the films exhibited a broad feature from 35°–37.5°, characteristic of amorphous NiO, which is common for NiO films deposited at room temperature.[24] As the temperature increased, the films crystallized into polycrystalline cubic NiO, consistent with prior observations.[24] The intensity of the (111) cubic NiO peak depended on sputtering pressure: the film deposited at 2.5 mTorr exhibited a relative peak intensity of 0.00425 at 1100 °C, compared with 0.05898 for the 5 mTorr film and 0.08123 for the 10 mTorr film. This pressure-dependent behavior is consistent with known effects of sputtering pressure on film morphology and microstructure.[25, 26]

The temperature-dependent 2θ peak position of the NiO (111) peak is shown in Fig. 1(d). The observed non-monotonic evolution of the peak position arises from a combination of amorphous-to-crystalline transition, reduction of point defects during thermal annealing, and thermal expansion at elevated temperatures.[27]

At a temperature of 1100 °C, figures 1(a)–1(c) reveal the emergence of an additional diffraction peak corresponding to the monoclinic $Ni_2O_3$ (110) reflection at a 2θ angle of 19.29°-19.38°. These findings suggest that the film initially transformed from amorphous NiOx, as shown in Fig. 2(a), to cubic NiO (Fig. 2(b)), at temperatures between 300 °C and 400 °C. The film then became a monoclinic $Ni_2O_3$ at 1100 °C (Fig. 2(c)). The $Ni_2O_3$ phase likely forms through oxidation of NiO. Among the three films studied, the 10 mTorr sample exhibited the weakest $Ni_2O_3$ peak at 1100 °C (relative intensity 0.003), followed by the 5 mTorr film (0.015) and the 2.5 mTorr film (relative intensity 0.024).

The sheet resistance of the films deposited at 2.5, 5.0, and 10 mTorr before annealing was 33, 6.0, and 68 kΩ/□, respectively. The sheet resistance of the films was too high to measure using a ST-

2258C four-probe tester post annealing. The increase in the sheet resistance was due to the decrease in the nickel vacancies[17] and possibly also due to the formation of $Ni_2O_3$.

Figure 3(a) plots the d-spacing of the cubic NiO (111) reflection with temperature, calculated using Bragg's law:[28]

$$n\lambda = 2d\sin(\theta) \tag{1}$$

Where $n$ is the diffraction order, $\lambda$ is the X-ray wavelength, $d$ is the interplanar spacing, and $\theta$ is the Bragg angle. The lattice constant $a$ of the NiO is presented in Fig. 3(b), and was obtained from the FCC relationship:

$$a = d\sqrt{3} \tag{2}$$

The lattice constant increased with temperature due to thermal expansion from 4.18, 4.19, and 4.18 Å at 300 °C to 4.19, 4.19, and 4.19 Å at 1100 °C for the films deposited at 2.5, 5, and 10 mTorr, respectively. The values obtained here deviated slightly from the reported lattice constants of powdered NiO,[29] which could be due to strain, defect density, and microstructure caused by the sputtering of the film. The films sputtered at different pressures all showed similar lattice constants at 1100 °C of 4.19 Å. The in-plane strain $\alpha$ was then calculated using (3), where $a_m$ is the measured lattice constant, and $a$ is the theoretical lattice constant. The films deposited at all pressures were in compressive strain. Previous studies have shown NiO films deposited using a PLD on sapphire were under tensile strain, which could be due to the variation in the deposition method [30].

$$\alpha = \frac{a-a_m}{a} \tag{3}$$

Figure 3(d) compares the unit cell volume as a function of temperature for the present films and reported values for powdered NiO.[29] The thermal expansion coefficient (TEC) for the NiO unit cell from 30 °C to 1000 °C is presented in Fig. 3(e). The TEC was calculated using (4), where $\frac{dV}{dT}$ represents the derivative of the unit cell volume with respect to temperature.

$$TEC = \frac{dV}{dT} \qquad (4)$$

The average grain size of the NiO films from 30 ˚C to 1000 ˚C is represented in Fig. 3(f), which was calculated using the Scherrer equation:[31]

$$D = \frac{k\lambda}{\beta cos(\theta)} \qquad (5)$$

Where $D$ is the average grain size, $k$ is the crystallite shape factor, and $\beta$ is the full-width at half-maximum of the NiO (111) peak. The grain size increased with temperature due to the continued crystallization of the film and was pressure-dependent, as the sputter pressure affected the microstructure of the deposited film.[25, 26]

The relative peak intensity of the NiO (111) film versus temperature is compared to the reported relative resistivity of NiO films, deposited at room temperature, and annealed at different temperatures,[21, 32, 33] shown by an Arrhenius plot in Fig. 4. The peak intensity and the conductivity were normalized to their initial value to facilitate the comparison. There is an increase in the slope in the conductivity of NiO films and the relative NiO (111) peak intensity between 200 ˚C to 300 ˚C. The slope of the peak intensity and the conductivity then reduce at ~400 ˚C. The increased resistivity and crystallinity may be a result of the elevated temperature, leading to the removal of Ni vacancies and the crystallization of NiO.[17] This suggests that devices that rely on NiO deposited at room temperature may be limited to operating temperatures of ~300 ˚C without risking severe increases in resistivity of the film in atmospheric environments. Packaging with a high partial pressure of $O_2$ could help increase the thermal equilibrium concentration of Ni vacancies in the film, thereby maintaining the film's stable resistivity. However, the increased presence of $O_2$ will result in the formation of $Ni_2O_3$ at elevated temperatures.

Figure 5 shows the XRD patterns of the films after the 1100 °C exposure, compared with the as-deposited film. The formation of the $Ni_2O_3$ phase occurred at 1100 °C, and remained present after cooling, indicating its stability. The relative intensity of the $Ni_2O_3$ peak followed the same pressure-dependent trend observed at 1100 °C.

In summary, sputtered NiO films deposited from 2.5–10 mTorr on c-plane sapphire were analyzed using *in situ* HT-XRD from 30 °C to 1100 °C. Initially amorphous films crystallized into cubic NiO between 300 °C and 400 °C and subsequently transformed into monoclinic $Ni_2O_3$ at a temperature of 1100 °C. Upon cooling, the $Ni_2O_3$ phase persisted, indicating irreversible oxidation. While all films underwent similar structural transitions, the relative intensities of both NiO and $Ni_2O_3$ reflections depended on sputtering pressure. Electrical testing revealed that the films became highly resistant after the 1100 °C exposure. The NiO (111) peak intensity followed a similar trend to the film resistivity, suggesting that elevated temperatures result in increased crystallinity and resistivity. These findings suggest that NiO films sputtered at room temperature may be limited to operating temperatures of ~300 ˚C without risking severe increases in resistivity of the film in atmospheric environments. Packaging that maintains a high $O_2$ partial pressure could help reduce the resistivity changes, but the increased presence of $O_2$ will result in the formation of $Ni_2O_3$ at elevated temperatures.

**AUTHOR DECLARATIONS**

**Conflict of Interest**

The authors have no conflicts to disclose.

**ACKNOWLEDGEMENT**

The authors acknowledge the support from the University of Utah start-up fund, PIVOT Energy Accelerator Grant U-7352FuEnergyAccelerator2023, and Pilot Funding through the HCI/Engineering Innovation in Cancer Engineering (ICE) Partnership Seed Grants. This work made use of the Nanofab EMSAL shared facilities of the Micron Technology Foundation Inc. Microscopy Suite, sponsored by the John and Marcia Price College of Engineering, Health Sciences Center, Office of the Vice President for Research. In addition, it utilized the University of Utah Nanofab shared facilities, which are supported in part by the MRSEC Program of the NSF under Award No. DMR-112125. Acquisition of the Bruker D8 Discover system was made possible by the Air Force Office of Scientific Research under project number FA9550-21-1-0293.

**Figures**

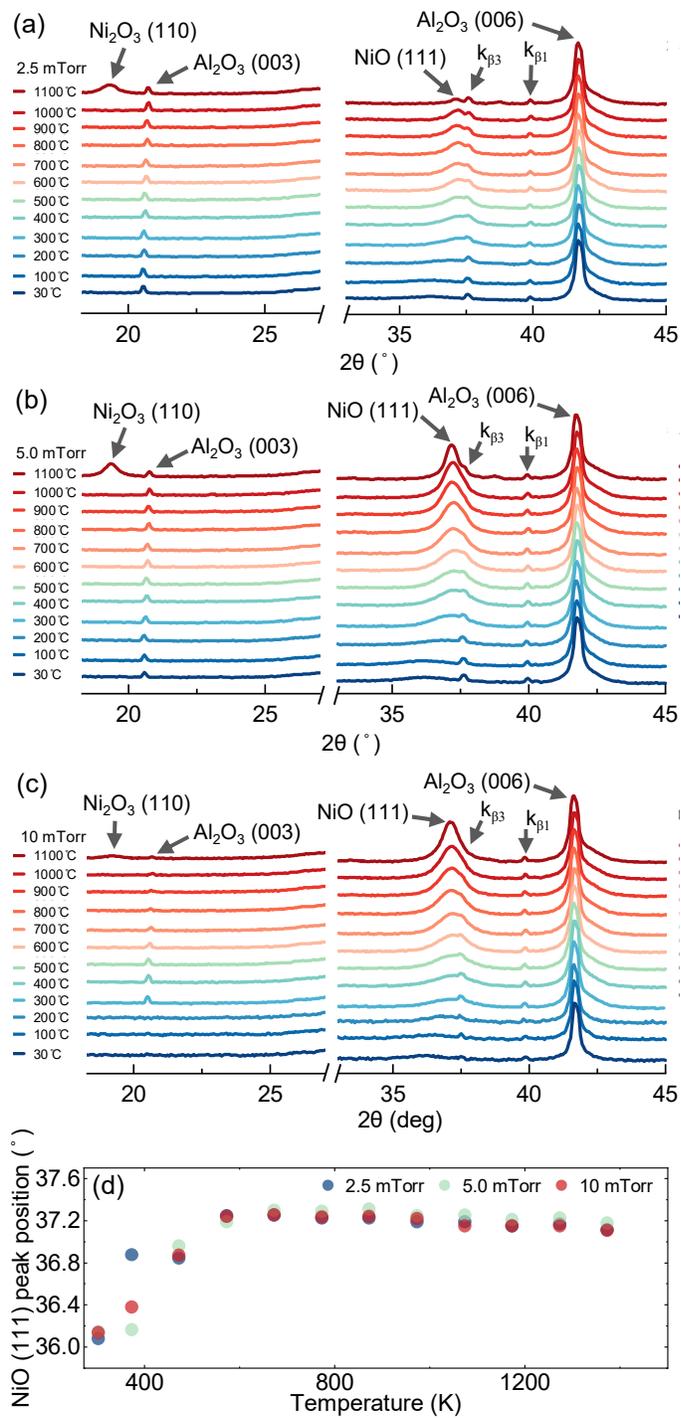

**Fig. 1.** *In situ* XRD patterns of the sputtered NiO film with a working pressure of (a) 2.5 mTorr, (b) 5.0 mTorr, and (c) 10 mTorr from 30 °C to 1100 °C. (d) 2θ peak position of the NiO (111) peak.

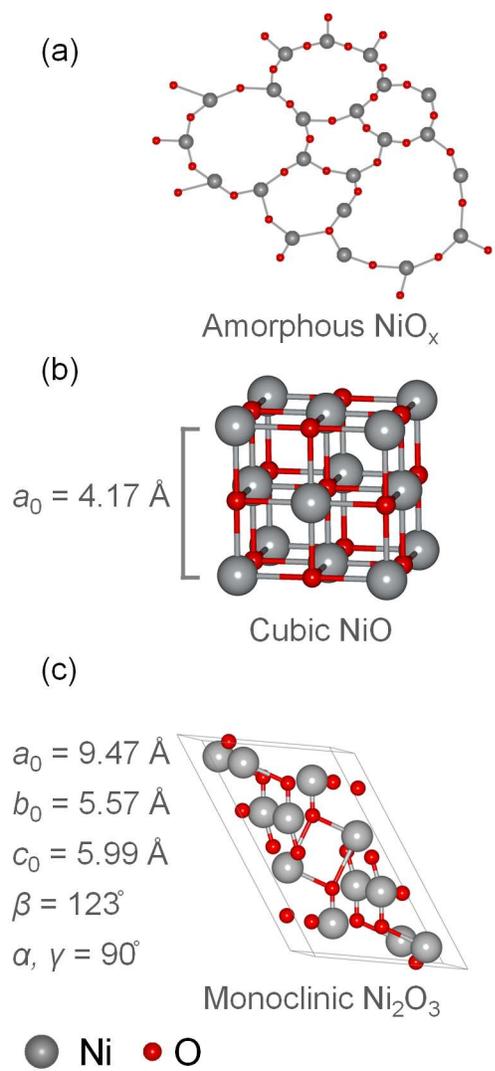

**Fig. 2.** Crystal structure model for (a) amorphous $NiO_x$, (b) cubic NiO, and (c) monoclinic $Ni_2O_3$.

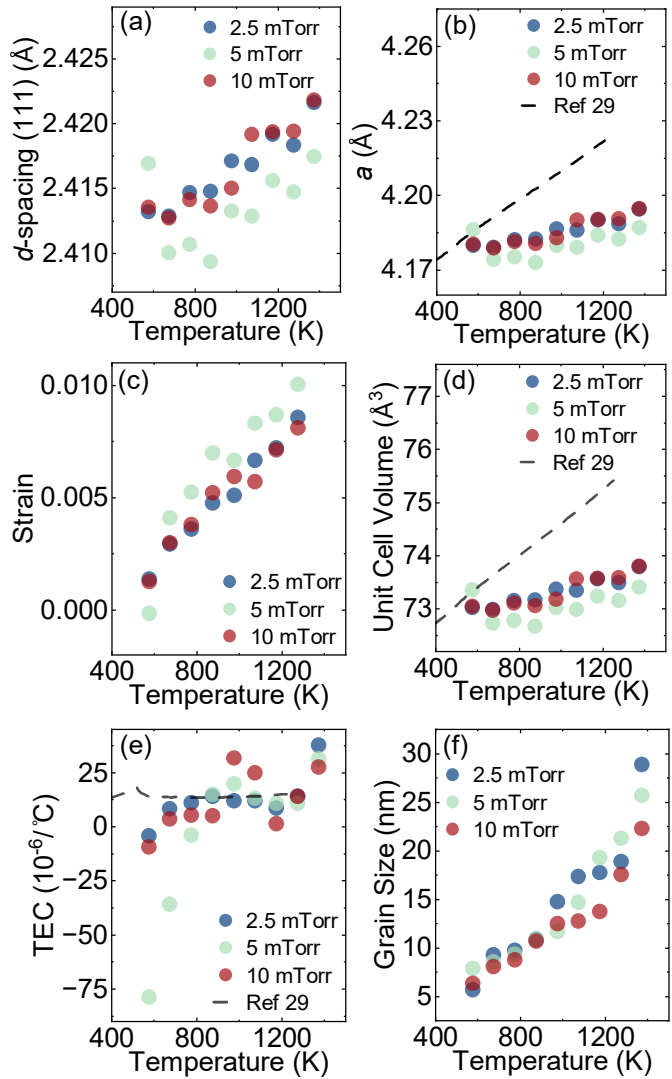

**Fig. 3.** (a) d-spacing along the (111) plane, (b) lattice constant (*a*), (c) strain in the NiO film, (d) unit cell volume, (e) thermal expansion coefficient (TEC), and (f) the average grain size of the cubic NiO.

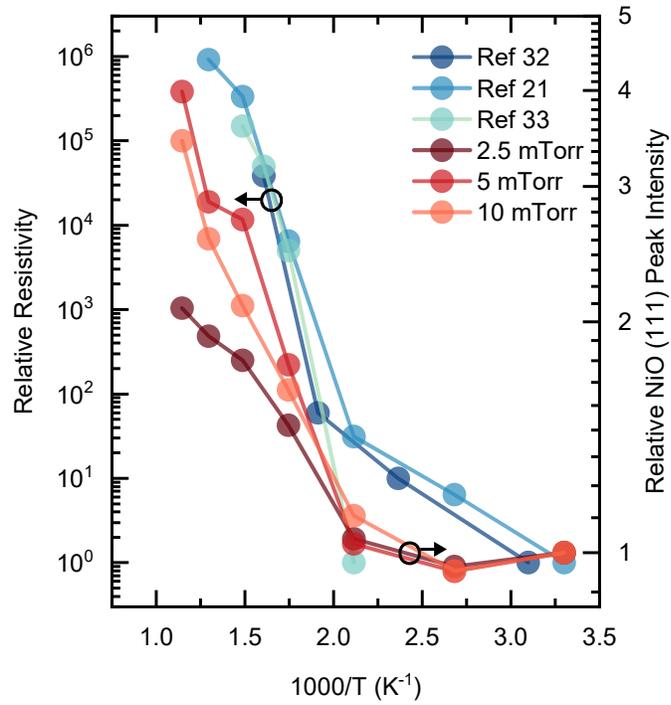

**Fig. 4.** Comparison of the relative resistivity of NiO films deposited at room temperature with the relative NiO (111) peak intensity sputtered at different pressures. The data was normalized to the initial value of the dataset.

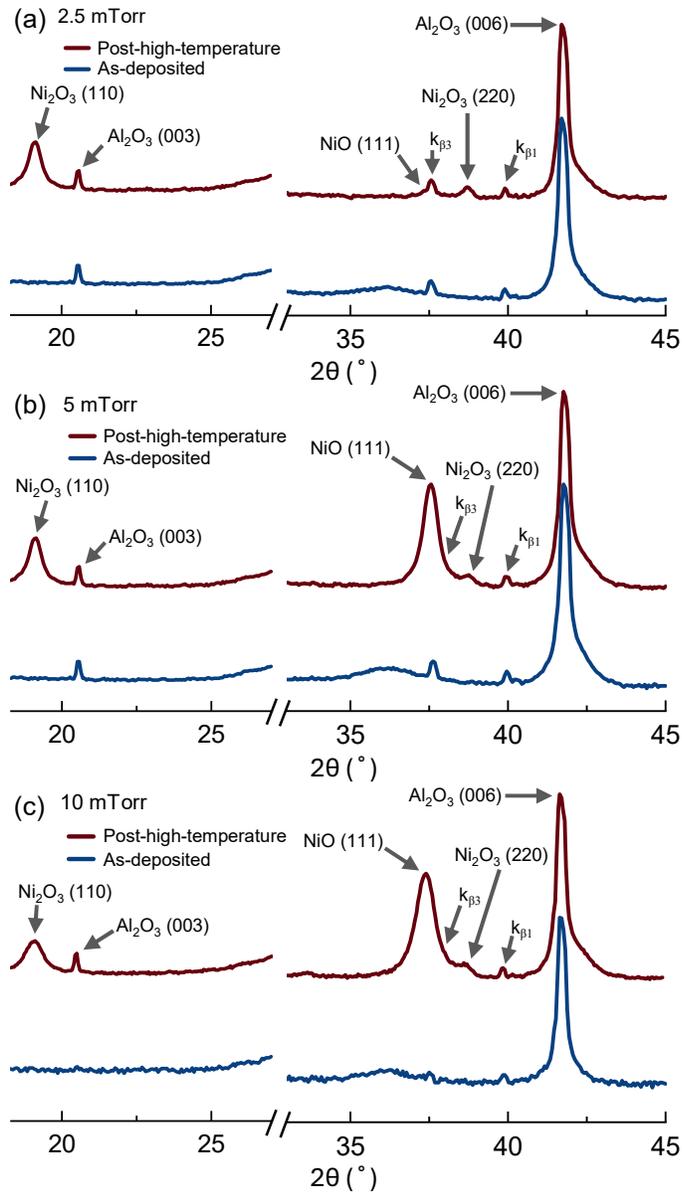

**Fig. 5.** Comparison of the XRD patterns before and after the high-temperature exposure at 30 °C with the film deposited at (a) 2.5 mTorr, (b) 5 mTorr, and (c) 10 mTorr.